\documentclass[reprint,superscriptaddress,pra,aps]{revtex4-2}
\usepackage{silence}
\usepackage[utf8]{inputenc}
\usepackage[english]{babel}
\usepackage{amsmath}
\usepackage{amssymb}
\usepackage{graphicx}
\usepackage{tikz}\usetikzlibrary{patterns}\usetikzlibrary{calc,3d}
\usepackage{stix}
\usepackage{placeins}
\usepackage{siunitx}
\usepackage{longtable}
\usepackage{siunitx}
\usepackage{microtype}
\usepackage{svg}\svgsetup{inkscapelatex=false,inkscapearea=page}
\usepackage{adjustbox} 
\usepackage[unicode=true,pdfusetitle,bookmarks=false,bookmarksnumbered=false,bookmarksopen=false,breaklinks=true,pdfborder={0 0 0},pdfborderstyle={},backref=false,colorlinks=false]{hyperref}
\usepackage{standalone}

\usepackage{natbib}
\usepackage{bibunits} 

\usepackage{array}
\newcolumntype{H}{>{\setbox0=\hbox\bgroup}c<{\egroup}@{}}

\usepackage[most]{tcolorbox}
\usepackage[normalem]{ulem}
\usepackage{xpatch}

\def\chanro#1{} 
\def\ccanc#1{} 
\def\green#1{} 
\def\gcanc#1{} 

\begin{document}
\title{Quadrature squeezing enhances Wigner negativity in a mechanical Duffing oscillator}

\author{Christian A. Rosiek}
\email{chanro@dtu.dk}
\altaffiliation{present address: DTU Electro, Department of Electrical and Photonics Engineering, Technical University of Denmark, {\O}rsteds Plads 343, Kgs.~Lyngby, DK-2800, Denmark}
\affiliation{Niels Bohr Institute, University of Copenhagen, Blegdamsvej 17, 2100, Copenhagen, Denmark}
\affiliation{Center for Hybrid Quantum Networks, Niels Bohr Institute, University of Copenhagen, Blegdamsvej 17, 2100, Copenhagen, Denmark}
\author{Massimiliano Rossi}
\altaffiliation{present address: Photonics Laboratory, ETH Z{\"u}rich, 8093 Z{\"u}rich, Switzerland}
\affiliation{Niels Bohr Institute, University of Copenhagen, Blegdamsvej 17, 2100, Copenhagen, Denmark}
\affiliation{Center for Hybrid Quantum Networks, Niels Bohr Institute, University of Copenhagen, Blegdamsvej 17, 2100, Copenhagen, Denmark}
\author{Albert Schliesser}
\affiliation{Niels Bohr Institute, University of Copenhagen, Blegdamsvej 17, 2100, Copenhagen, Denmark}
\affiliation{Center for Hybrid Quantum Networks, Niels Bohr Institute, University of Copenhagen, Blegdamsvej 17, 2100, Copenhagen, Denmark}
\author{Anders S. S{\o}rensen}
\email{anders.sorensen@nbi.ku.dk}
\affiliation{Niels Bohr Institute, University of Copenhagen, Blegdamsvej 17, 2100, Copenhagen, Denmark}
\affiliation{Center for Hybrid Quantum Networks, Niels Bohr Institute, University of Copenhagen, Blegdamsvej 17, 2100, Copenhagen, Denmark}

\begin{abstract}
Generating macroscopic non-classical quantum states is a long-standing challenge in physics. Anharmonic dynamics is an essential ingredient to generate these states, but for large mechanical systems, the effect of the anharmonicity tends to become negligible compared to decoherence. As a possible solution to this challenge, we propose to use a motional squeezed state as a resource to effectively enhance the anharmonicity. We analyze the production of negativity in the Wigner distribution of a quantum anharmonic resonator initially in a squeezed state. We find that initial squeezing enhances the rate at which negativity is generated. We also analyze the effect of two common sources of decoherence, namely energy damping and dephasing, and find that the detrimental effects of energy damping are suppressed by strong squeezing. In the limit of large squeezing, which is needed for state-of-the-art systems, we find good approximations for the Wigner function. Our analysis is significant for current experiments attempting to prepare macroscopic mechanical systems in genuine quantum states. We provide an overview of several experimental platforms featuring nonlinear behaviors and low levels of decoherence. In particular, we discuss the feasibility of our proposal with carbon nanotubes and levitated nanoparticles.
\end{abstract}
\maketitle

\let\oldaddcontentsline\addcontentsline
\def\addcontentsline#1#2#3{}

\section{Introduction}
Quantum theory has revolutionized our understanding of the microscopic world.
As of today, however, quintessential quantum phenomena remain elusive in
large-scale systems. 
One possible explanation is the breakdown of quantum mechanics at large scales, as predicted by collapse models~\cite{ghirardi_unified_1986, bassi_models_2013}.
Bringing macroscopic systems to the quantum regime thus has a notable interest in quantum science.
A unique feature of non-classical quantum states is a region of negative values in their Wigner phase space distribution, termed Wigner negativity~\cite{kenfack_negativity_2004, zurek_sub-planck_2001}.
Current research is attempting to generate and observe this Wigner negativity in the motion of mechanical systems of ever larger size.
Wigner negativity has been observed, so far, in microscopic
systems, either comprising only up to a few thousand elementary constituents or
possessing a wavefunction spatially localized to sub-atomic distances~\cite{leibfried_experimental_1996, hofheinz_synthesizing_2009, chu_creation_2018, fein_quantum_2019, bild_schrodinger_2023}.

One of the challenges opposing the generation of non-classical states is to avoid decoherence, that is the loss of quantum coherence due to the unwanted interaction with an environment (comprising, e.g., air molecules, acoustic phonons, and thermal photons) \cite{schlosshauer_decoherence_2007}.
Recently, researchers have developed novel techniques that have enabled excellent isolation of macroscopic mechanical resonators from their environment, thereby minimizing decoherence~\cite{MacCabe_NanoacousticResonatorUltralong_2020,tsaturyan_ultracoherent_2017}.
Simultaneously, the
field of optomechanics has refined paradigms to control the mechanical motion at
the level of single quanta. 
These efforts culminated in ground-state cooling of a variety of mechanical
systems~\cite{chan_laser_2011,oconnell_quantum_2010,Rossi_MeasurementbasedQuantumControl_2018,tebbenjohanns_quantum_2021,Delic_CoolingLevitatedNanoparticle_2020}.
These states are, however, regarded as semiclassical because they have Gaussian Wigner functions, which have no negativity.

To go beyond Gaussian physics requires anharmonic dynamics, such as those generated by a nonlinear crystal in quantum optics \cite{hong_experimental_1986}, by Josephson junctions in superconducting circuits \cite{grimm_stabilization_2020} or by a diffraction grating in matter-wave interferometry \cite{hornberger_colloquium_2012}.
In the context of mechanical resonators, the anharmonicity usually comes from the nonlinear contributions to the restoring force. Such contributions are, in general, small so throughout this paper we retain only the first correction, which scales cubically with the displacement from the equilibrium. This nonlinear force is known as Duffing nonlinearity and, in the absence of decoherence, is sufficient to generate negativity in a driven system \cite{stobinska_generation_2011}. 
The typical signature of this anharmonicity is an amplitude-dependent resonance frequency \cite{schmid_fundamentals_2016}.
Roughly speaking, the quantum effects of this nonlinearity become significant when the frequency shift induced by the typical quantum motion (e.g., the zero-point fluctuations for a resonator in its ground state) exceeds the decoherence rate. In practice, no existing mechanical platforms meet this requirement and, thus, generating negativity via nonlinearity remains a daunting task.\\
\indent There are two paths to overcome this limitation: engineering systems with larger anharmonicity or increasing the size of the typical quantum motion. In this work, we propose and analyze the latter alternative.
The position of a mechanical resonator in the ground state is not well determined but has residual fluctuations as required by the Heisenberg uncertainty principle. One can enlarge these fluctuations while simultaneously reducing, or squeezing, the fluctuations in momentum.
We consider such a squeezed state as the initial state of a Duffing resonator. The increased position fluctuations generate both a larger Duffing frequency shift and a larger decoherence rate. Nevertheless, we show that the generated negativity scales more favorably with the initial squeezing than the decoherence. Finally, we present an overview of existing Duffing systems and analyze the feasibility of our proposal with carbon nanotubes and levitated nanoparticles.

\section{Unitary anharmonic dynamics}
\begin{figure}
  \includegraphics[width=8.6cm]{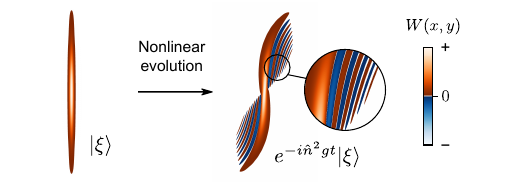}
  \caption{Protocol concept.
  A system is prepared in an initial squeezed vacuum state, $|\xi \rangle$. Subsequently, the system evolves in a Duffing potential. The anharmonicity bends the state's Wigner function $W$ and generates fringes and negativity. Regions where $W$ is negative are shown as blue.
  }
  \label{fig:concept-sketch}
\end{figure}

We first analyze the unitary dynamics of an anharmonic mechanical resonator with a Duffing nonlinearity.
The respective Hamiltonian is
\begin{equation}
  \hat H = {\hat p^2\over 2m} + \frac{1}{2}m\Omega_m^2\hat q^2 + {\beta\over4}\hat q^4\,,
\end{equation}
where $\hat q$ and $\hat p$ describe the position and momentum of the mechanical resonator, $m$ denotes its mass, $\Omega_m$ denotes its resonance frequency, and $\beta$ is the Duffing parameter which describes the change in stiffness for a given displacement amplitude.
By changing to a rotating frame with a frequency $\omega_0\approx \Omega_m$ and discarding fast-oscillating terms (see Supplemental Material), we arrive at the simplified Hamiltonian
\begin{equation}
  \label{eq:rwa-hamiltonian}
  \hat H = \hbar g (\hat n^2 - \hat n) \,,
\end{equation}
where $g = 3\hbar\beta/8m^2\Omega_m^2$ is a parameter quantifying the nonlinear coupling, $\hat{n}=\hat{q}^2/(4q_\mathrm{zpf}^2)+\hat{p}^2/(4p_\mathrm{zpf}^2)-1/2$ is the phonon number operator and $q_\mathrm{zpf}=\sqrt{\hbar/(2 m \Omega_m)}$, $p_\mathrm{zpf}=\sqrt{\hbar m \Omega_m/2}$ are position and momentum zero-point fluctuations, respectively.
The Hamiltonian in Eq.~\eqref{eq:rwa-hamiltonian}, despite its simple appearance, describes a nonlinearity that is able to generate non-classical correlations for a quantum system and thereby introduce negative values in the Wigner function if given an appropriate initial state, as sketched in Fig.~\ref{fig:concept-sketch} \cite{Stobinska_WignerFunctionEvolution_2008}.
The rotating wave approximation made above is valid when $g\ll\Omega_m$, which is an excellent approximation for most relevant optomechanical systems.
In the following, we work with the dimensionless quadrature operators in the interaction picture, defined as $\hat{X} = (2q_{\mathrm{zpf}})^{-1}e^{i\hbar\Omega_m \hat n} \,\hat q \,e^{-i\hbar\Omega_m \hat n}$ and $\hat{Y} = (2p_{\mathrm{zpf}})^{-1}e^{i\hbar\Omega_m \hat n}\hat p\,e^{-i\hbar\Omega_m \hat n}$ and satisfying the commutation relation $[\hat{X},\hat{Y}]=i/2$. 
To characterize the property of the state, we compute the Wigner quasiprobability distribution in the same interaction picture. We label with $x$ and $y$ the coordinates corresponding to the quadratures $\hat{X}$ and $\hat{Y}$, respectively.

The vacuum state is an eigenstate of Eq.~\eqref{eq:rwa-hamiltonian}, thus a system prepared in that state will not show any negativity during the evolution.
Instead, a coherent state obtained by displacing the vacuum state will develop fringes with negative areas in the Wigner function.
However, this process is highly sensitive to decoherence, posing a limit to the amount of displacement that can be employed \cite{Stobinska_WignerFunctionEvolution_2008}.
To overcome this limitation, we study the dynamics for a different initial state.
In particular, we focus our attention on squeezed vacuum states and show that they are more robust against decoherence.

\begin{figure}
  \includegraphics[width=8.6cm]{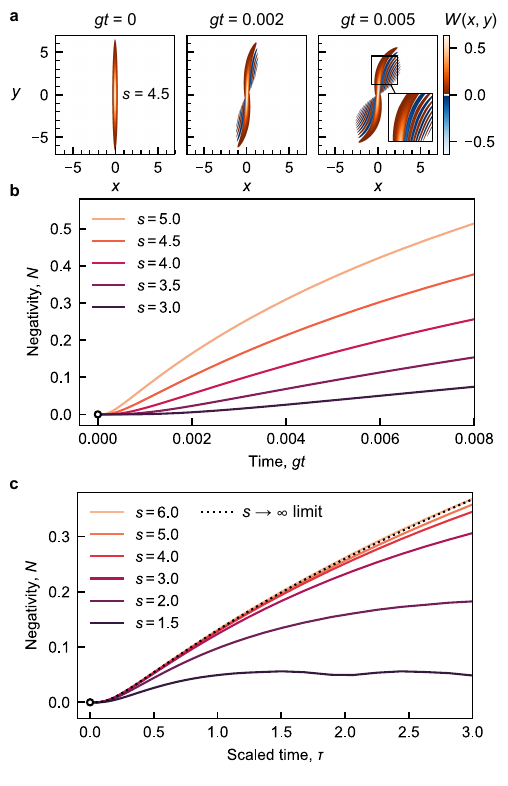}
  \caption{Unitary evolution of the Wigner function.
    \textbf{a}, Phase space plot of the Wigner function, $W(x,y)$, of an initial squeezed state with $s=4.5$ (left panel) and phase space plots showing the S-shape formed by unitary evolution under the Hamiltonian of Eq.~(\ref{eq:rwa-hamiltonian}) at times $gt=0.002$ (center panel) and $gt=0.005$ (right panel). This corresponds to the scaled times $\tau =gts^4=0.82$ and $\tau=2.05$. The inset on the right shows the details of the negative fringes of $W(x,y)$, the integral of which is denoted the negativity, $N$.
  \textbf{b}, Time evolution of negativity, $N$, for different initial degrees of initial squeezing.
  \textbf{c}, Time evolution of $N$ as in \textbf{b}, but with a rescaled time axis, $\tau = gts^4$.
  The dotted black line shows the negativity obtained from the solution in Eq.~\eqref{eq:unitary-fourier-solution} corresponding to the limit of infinite squeezing, $s\to\infty$.}
  \label{fig:initial-evolution}
\end{figure}

The Wigner function of such a state, squeezed along the $x$-quadrature, is
\begin{equation}
  W(x,y,0) = {2\over\pi}e^{-2x^2s^2-2y^2/s^2}\,,
\end{equation}
where $s$ quantifies the amount of squeezing and the quadrature variances are given by $\langle \hat{X}^2\rangle - \langle \hat{X}\rangle^2 = (s/2)^{-2}$ and $\langle \hat{Y}^2\rangle - \langle \hat{Y}\rangle^2 = (s/2)^2$, where $\langle\cdot \rangle$ indicates a quantum expectation value.
Due to the rotational symmetry in phase space, we can assume $s\ge1$ without loss of generality.
In Fig.~\ref{fig:initial-evolution}a, we show the initial state with $s=4.5$.
We compute, in a Fock state basis, the time-evolved state, from which we then calculate the Wigner function using the software QuTiP \cite{johansson_qutip_2013}. We also show these states in Fig.~\ref{fig:initial-evolution}a, for the chosen times $gt=0.002$ and $gt=0.005$.
We notice that the Wigner function exhibits characteristic fringes with negative values during this short time period of the evolution.
This is the quantum signature we seek \cite{kenfack_negativity_2004}.

While Eq.~\eqref{eq:rwa-hamiltonian} is easily solvable in the number state basis, we can gain more insight about the appearance of negative values in the Wigner function by studying the equation of motion for the Wigner function instead.
The evolution of the Wigner function is described by a partial differential equation (PDE) derived from the Hamiltonian
\cite{Walls_QuantumOptics_2008}. The PDE corresponding to Eq.~\eqref{eq:rwa-hamiltonian} is
\begin{equation}
  \label{eq:unitary-eom}
  \begin{split}
    \partial_t W(x,y,t) &= 2g(x^2+y^2)\left(-y\partial_x + x\partial_y\right)W(x,y,t) \\
    &\phantom{=} +{g\over8}\left(-y\partial_x^3 + x\partial_y^3 + x\partial_y\partial_x^2 - y\partial_x\partial_y^2\right)W(x,y,t) \,.
  \end{split}
\end{equation}
The first line of Eq.~\eqref{eq:unitary-eom} describes a radially varying phase space rotation with an angular rotation proportional to $x^2+y^2$, leading to a shearing of the Wigner function around the origin (the S-shape visible in Fig.~\ref{fig:initial-evolution}a).
This shearing is also present in the classical Duffing oscillator, and cannot by itself produce a negative Wigner function.
The classical nature of the first line in Eq.~\eqref{eq:unitary-eom} can be highlighted by considering the limit $\hbar\to 0$.
According to their definitions, the nonlinear coupling parameter $g$ is proportional to $\hbar$, whereas the coordinates $x$ and $y$ are both proportional to $\hbar^{-1/2}$. 
Thus, only terms in the second line are proportional to $\hbar$ and Eq.~\eqref{eq:unitary-eom} reduces to the Fokker-Planck equation for a classical nonlinear oscillator in the limit $\hbar\to0$.

Terms in the second line of Eq.~\eqref{eq:unitary-eom} describe a non-classical effect, which can lead to negative values of $W(x,y,t)$ after finite evolution.
We quantify the degree of non-classicality of the oscillator state by calculating the volume of the negative part of the Wigner function, i.e.,
\begin{equation}
  N = -\iint dx\,dy\,\min\{W(x,y),0\},
\end{equation}
where the integration is carried out in the whole phase space.
This figure of merit has already found use in previous works \cite{kenfack_negativity_2004,Arkhipov_NegativityVolumeGeneralized_2018}. For brevity, we simply refer to it as ``negativity''.
For instance, for Gaussian states we find $N=0$. In the sense of this definition, Gaussian states are classical states~\cite{hudson_when_1974}.
The evolution of $N$ depends on the squeezing of the initial state. Figure~\ref{fig:initial-evolution}b shows the initial evolution of $N$ for squeezed initial states with varying squeezing parameters, $s$. From $N=0$ for the initial state, the initial evolution has $N$ growing as the fringes form. Notably, initial states with more squeezing generally exhibit a faster increase in $N$.

To understand the dynamics leading to the formation of negativity, we make some approximations to Eq.~\eqref{eq:unitary-eom}.
We introduce a set of rescaled coordinates $\tilde x=sx$ and $\tilde y = y/s$.
The rescaled Wigner function is $\tilde W(\tilde x, \tilde y, t) = W(\tilde x/s, s\tilde y, t)$, and the rescaled differential operators $\partial_{\tilde x} = (1/s)\partial_x$ and $\partial_{\tilde y} = s\partial_y$. 
Making the substitutions in Eq.~\eqref{eq:unitary-eom}, we arrive at
\begin{equation}
  \label{eq:90}
  \begin{split}
    & \partial_t \tilde W(\tilde x,\tilde y,t) \\
    &= 2g
    \left(
    -\tilde x^2\tilde y\partial_{\tilde x} - s^4\tilde y^3\partial_{\tilde x} + {1\over s^4}\tilde x^3\partial_{\tilde y} + \tilde x\tilde y^2\partial_{\tilde y}
    \right)
    \tilde W(\tilde x,\tilde y,t) \\
    &\mathrel{\phantom{=}} +{g\over8}
    \left(
    -s^4\tilde y\partial_{\tilde x}^3 + {1\over s^4}\tilde x\partial_{\tilde y}^3 + \tilde x\partial_{\tilde y}\partial_{\tilde x}^2 - \tilde y\partial_{\tilde x}\partial_{\tilde y}^2
    \right)
    \tilde W(\tilde x,\tilde y,t)\,.
  \end{split}
\end{equation}
In terms of these rescaled coordinates, the initial state Wigner function has no dependence on the squeezing parameter, $s$, and becomes symmetric in $\tilde x$ and $\tilde y$, taking the form $\tilde W(\tilde x,\tilde y,0)~=~(2/\pi)\exp\big(-2(\tilde x^2+\tilde y^2)\big)$.
In other words, we have moved the explicit dependence on initial squeezing from the state to the system dynamics.
In many disparate systems, the nonlinear coupling parameter $g$ is weak (see discussions in Sec.~\ref{sec:conclusions}).
We are interested in the regime of large initial squeezing, to eventually enhance this rate.
Assuming $s\gg1$ and noticing that the coordinates $\tilde{x}$ and $\tilde{y}$ are $\mathcal{O}(1)$, at least during the initial evolution, we retain only terms proportional to $s^4$ in Eq.~\eqref{eq:90}.
The resulting approximated equation is
\begin{equation}
  \label{eq:large-squeezing-eom}
  \partial_t \tilde W(\tilde x,\tilde y,t)
  \approx -2gs^4\tilde y^3\partial_{\tilde x}\tilde W(\tilde x,\tilde y,t)
  -{gs^4\over8}\tilde y\partial_{\tilde x}^3\tilde W(\tilde x,\tilde y,t)\,.
\end{equation}
We note that all right-hand side terms have as a common factor the frequency $gs^4$.
This parameter dictates the rate at which the system evolves and, especially, the rate at which negativity builds up.
Figure~\ref{fig:initial-evolution}c shows the negativity $N$ as a function of the rescaled time $\tau = gts^4$.
In these units, all the curves overlap for large squeezing and for short time, suggesting that the negativity indeed depends on the rescaled time only.
We can introduce a parameter-free model that captures this universal behavior.
We notice that Eq.~\eqref{eq:large-squeezing-eom} has been reduced to a PDE in $\tilde x$ and $t$ only. 
For a given $\tilde y$ coordinate, the solution can be expressed in terms of inverse Fourier transform as
\begin{equation}
  \label{eq:unitary-fourier-solution}
  \tilde W(\tilde x,\tilde y,\tau) = {1\over\sqrt{2\pi}}\int_{-\infty}^\infty dk \, \tilde h_{\tilde y}(k) \, e^{-2k\tau\tilde y^3 - \tau k^3\tilde y/8}e^{ik\tilde x},
\end{equation}
where
\begin{equation}
  \label{eq:initial-state-fourier-transform}
  \tilde h_{\tilde y}(k) = {1\over\sqrt{2\pi}} \int_{-\infty}^\infty d\tilde x \tilde W(\tilde x,\tilde y,0)e^{-ik\tilde x}\,
\end{equation}
is the Fourier transform of the initial state at a given $\tilde y$-coordinate.
The value of $N$ calculated from this approximate solution \eqref{eq:unitary-fourier-solution} is overlayed with the full solutions in Fig.~\ref{fig:initial-evolution}c.
The good agreement with the numerical solutions shows the validity of our approximations. 
In particular, it shows that the nonlinear coupling parameter $g$ is effectively enhanced by the factor $s^4$, the main finding of our analysis.
This represents a favorable scaling especially when compared to one of the decoherence processes, as we will show in the next section.

\section{Effects of decoherence}
In the previous section, we discussed the ideal case of purely unitary evolution of an anharmonic oscillator.
In practice, however, every system is coupled to some sort of environment which causes decoherence of the
state.
The decoherence usually competes with the generation of negativity and, eventually, gives rise to a semiclassical steady state.
In addition, decoherence affects different initial states differently, e.g., it is faster for a squeezed state compared to a coherent state \cite{schlosshauer_decoherence_2007}.
We augment the model previously introduced with two common decoherence mechanisms, namely energy damping and dephasing.
We show that, in some cases, the use of squeezing to enhance the nonlinearity is capable of outpacing the increased decoherence rates that accompany squeezed states.

\begin{figure}
  \includegraphics[width=8.6cm]{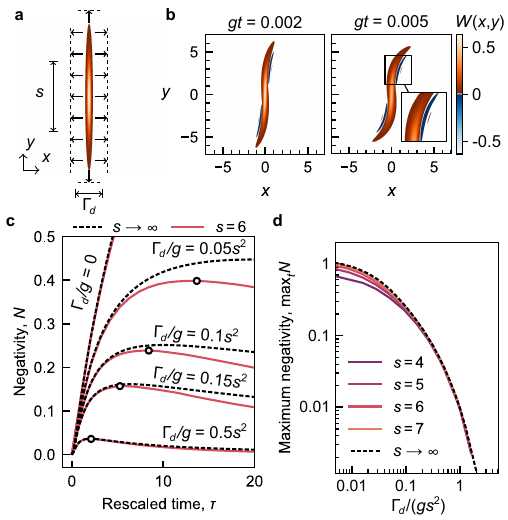}
  \caption{Generation of negativity in the presence of damping.
    \textbf{a}, Pictorial illustration of the effect of damping on a squeezed state.
    Even though the heating rate is homogeneous in both directions, damping causes a more pronounced broadening of the state along $x$, where the state varies more rapidly.
    \textbf{b}, Phase space plots of the Wigner function for $s=4.5$ evolved with a decoherence rate of $\Gamma_{\mathrm{d}}/g=0.5s^2$ (with $\bar{n}_{\mathrm{th}}=1000$). The initial state and shown times are shared with Fig.~\ref{fig:initial-evolution}a.
    Damping reduces the fringes which results in a smaller negativity, $N$ (compare to Fig.~\ref{fig:initial-evolution}a).
    \textbf{c}, Time evolution of negativity for initial squeezed state with $s=6$ as a function of rescaled time $\tau$ for various damping rates, $\Gamma_{\mathrm{d}}$, between $0$ and $0.5s^2/g$. The damping limits the negativity, with increased damping causing the maximum negativity, $\max_tN$, to peak earlier and lower. The peak negativity is indicated by a circle for each damped graph. The dashed line shows approximate analytical solution given in Eq.~\eqref{eq:damping-fourier-solution} corresponding to the limit of large squeezing, $s\to\infty$.
    \textbf{d}, Maximum negativity achieved, $\max_tN$, as a function of the ratio between the enhanced decoherence and nonlinear coupling rates, $\Gamma_{\mathrm{d}} s^2$ and $g s^4$, respectively for four finite squeezings, $s$ (see legend). The maximum negativity in the limit of large squeezing, obtained as in \textbf{c}, is shown as a dashed line. The maximum achieved negativity approaches this limit as the squeezing increases.
    \label{fig:damping}}
\end{figure}

We start by introducing decoherence via energy damping.
This damping arises from a linear coupling between the oscillator and a thermal bath, which is described by a mean phonon occupancy $\bar{n}_\mathrm{th}$ and a coupling rate $\gamma$ \cite{giovannetti_phase-noise_2001}.
The equation of motion for the Wigner function, including damping, is~\cite{Stobinska_WignerFunctionEvolution_2008}
\begin{equation}
  \label{eq:damping-eom}
  \begin{split}
    \partial_t W(x,y,t)
    &= LW(x,y,t) + {\Gamma_{\mathrm{d}}\over4}\nabla^2W(x,y,t) \\
    &\mathrel{\phantom{=}} + {\gamma\over2}\partial_x\left(xW(x,y,t)\right) + {\gamma\over2}\partial_y\left(yW(x,y,t)\right)
  \end{split}
\end{equation}
where the first term, $LW(x,y,t)$, describes the unitary evolution, i.e., the right-hand side of Eq.~\eqref{eq:unitary-eom}, the second term, proportional to the decoherence rate, $\Gamma_{\mathrm{d}}=\gamma(\bar{n}_\mathrm{th} + 1/2)$, describes a diffusion and the last two terms concentrate the Wigner distribution toward the origin.
This damping tends to reduce the negativity, $N$, and, e.g., systems evolved under damping only, i.e., where $LW(x,y,t)=0$ in Eq.~\eqref{eq:damping-eom}, will in general evolve to have $N=0$ within a finite time (see Supplemental Material).
For high temperatures and short times, i.e., $\bar{n}_\mathrm{th} \gg 1$ and $t \ll 1/\Gamma_{\mathrm{d}}$,
we may neglect the last two terms of Eq.~(\ref{eq:damping-eom}) and only retain the diffusive term for the damping.
This diffusion is uniform in the entire phase space, i.e., the coefficient of the second term in Eq.~\eqref{eq:damping-eom} is constant, as illustrated in Fig.~\ref{fig:damping}a.
We numerically solve the master equation corresponding to Eq.~\eqref{eq:damping-eom} (see Supplemental Material for details) and show examples of computed Wigner functions during the evolution in Fig.~\ref{fig:damping}b.
Comparing these plots to the ones with no damping (cf. Fig.~\ref{fig:initial-evolution}a), we notice that fringes and negativity are still present, but are reduced in magnitude.
In Fig.~\ref{fig:damping}c, we plot the evolution in units of rescaled time of the negativity, for different decoherence rates.
The diffusion causes the fringes to die out, eventually eliminating all negativity.
As expected, this effect is more pronounced and its onset is earlier for
higher decoherence rate $\Gamma_{\mathrm{d}}$.
We rescale now the coordinates in the same manner as shown in the previous section.
The transformed equation is cumbersome and does not yield much further intuition so we only provide it in the Supplemental Material for completeness.
Here, we rather focus on the limit of large initial squeezing, $s\gg1$.
In this case, we retain only the terms in the highest power of $s$ and arrive at
\begin{equation}
  \label{eq:large-squeezing-damping-eom}
  \begin{split}
  \partial_t \tilde W(\tilde x,\tilde y,t)
  &\approx -2gs^4\tilde y^3\partial_{\tilde x}\tilde W(\tilde x,\tilde y,t)
  -{gs^4\over8}\tilde y\partial_{\tilde x}^3\tilde W(\tilde x,\tilde y,t) \\
  &\mathrel{\phantom{=}} + {\Gamma_{\mathrm{d}}s^2\over4}\partial_{\tilde x}^2\tilde W(\tilde x, \tilde y, t)\,.
  \end{split}
\end{equation}
The system now has two typical timescales:
negativity is produced at the rate $gs^4$ while energy damping destroyes the negativity at the rate $\Gamma_{\mathrm{d}} s^2$.
In particular, we highlight that increasing the initial squeezing enhances the nonlinearity more than the energy damping, leading to an effectively larger generation of negativity.
This is one of the main results of this work.

To further substantiate this result, we show in Fig.~\ref{fig:damping}d the maximum negativity obtained during the time evolution (cf.  Fig.~\ref{fig:damping}c) as a function of the dimensionless parameter $\Gamma_{\mathrm{d}}/(gs^2)$.
Larger negativity is obtained for small decoherence rate $\Gamma_{\mathrm{d}}$ and for large squeezing $s$.
Finally, we can write the following explicit form of the solution to Eq.~\eqref{eq:large-squeezing-damping-eom} using a Fourier transformation:
\begin{equation}
  \label{eq:damping-fourier-solution}
  \tilde W(\tilde x,\tilde y,t) = {1\over\sqrt{2\pi}}\int_{-\infty}^\infty dk \, \tilde h_{\tilde y}(k) \,
  \begin{aligned}[t]
     &e^{ik\tilde x-2kgts^4\tilde y^3 - gts^4k^3\tilde y/8} \\
     &\times e^{-\Gamma_{\mathrm{d}} s^2t/8}
  \end{aligned}
\end{equation}
with $\tilde h_{\tilde y}(k)$ still defined by Eq.~(\ref{eq:initial-state-fourier-transform}).
To demonstrate the applicability of the approximation, we compute the maximum negativity based on Eq.~\eqref{eq:damping-fourier-solution} as we vary the parameter $\Gamma_{\mathrm{d}}/(gs^2)$. We show the result in Fig.~\ref{fig:damping}d and observe a good agreement with the full numerical analysis.

\begin{figure}
  \includegraphics[width=8.6cm]{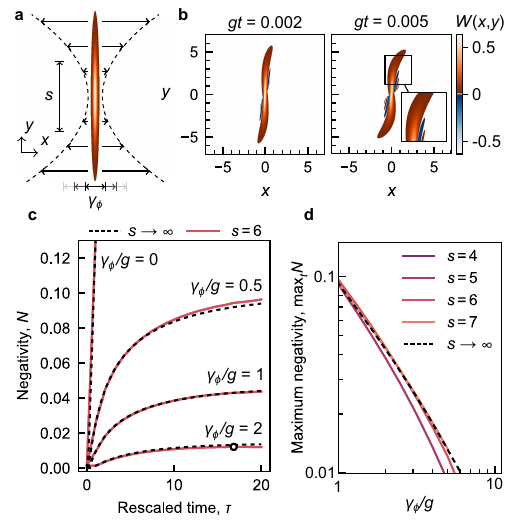}
  \caption{Generation of negativity in the presence of dephasing.
    \textbf{a}, Pictorial illustration of the effect of dephasing on a squeezed state. Compared with linear damping (Fig.~\ref{fig:damping}a), the diffusive effect of dephasing increases with distance to the origin.
    \textbf{b}, Phase space plots of the Wigner function for $s=4.5$ evolved with a dephasing rate of $\gamma_\phi/g=1$. The initial state and shown times are shared with Fig.~\ref{fig:initial-evolution}a. Like damping, dephasing reduces the fringes, though dephasing has a smaller effect on fringes closer to the origin.
    \textbf{c}, Time evolution of negativity for different initial degrees of squeezing in the presence of dephasing, as a function of rescaled time $\tau$.
    \textbf{d}, Maximum values of negativity from \textbf{c} as a function of the ratio between the enhanced dephasing and nonlinear coupling rates, $\gamma_\phi s^4$ and $g s^4$, respectively.
    \label{fig:dephasing}
  }
\end{figure}

Energy damping is only one of the possible sources of decoherence in the experimental reality.
Another important source is represented by drifts and fluctuations of the oscillator resonance frequency \cite{sansa_frequency_2016}.
We model the associated decoherence, sometimes referred to as dephasing, with the following equations of motion
\begin{equation}
  \label{eq:dephasing-eom}
  \begin{split}
    \partial_t W(x,y,t)
    &= LW(x,y,t) + {\gamma_\phi\over 2}(-y\partial_x + x\partial_y)^2W(x,y,t)\,,
  \end{split}
\end{equation}
where the first term, $LW(x,y,t)$, is again the unitary evolution as in Eq.~(\ref{eq:damping-eom}) and $\gamma_\phi$ is the dephasing rate.
The derivation is reported in the Supplemental Material.
The dynamics generated by dephasing mathematically corresponds to a diffusion in the azimuthal coordinate.
Effectively, this means that the effect of dephasing is stronger in the regions of phase space that are farther from the origin, as is illustrated in Fig.~\ref{fig:dephasing}a (compare with Fig.~\ref{fig:damping}a).
This has the effect of concentrating the negativity toward the phase space origin as shown in Fig.~\ref{fig:dephasing}b.
Figure~\ref{fig:dephasing}c shows the initial evolution of negativity under varying rates of dephasing, $\gamma_\phi$.
We can analyze this effect by transforming to the rescaled coordinates and dropping all terms except for the leading terms in $s$. In this way, we approximate Eq.~\eqref{eq:dephasing-eom} with
\begin{equation}
  \label{eq:large-squeezing-dephasing-eom}
  \begin{split}
    \partial_t \tilde W(\tilde x,\tilde y,t)
    &\approx -2gs^4\tilde y^3\partial_{\tilde x}\tilde W(\tilde x,\tilde y,t)
  -{gs^4\over8}\tilde y\partial_{\tilde x}^3\tilde W(\tilde x,\tilde y,t) \\
    &\mathrel{\phantom{=}} + {\gamma_\phi s^4\over 2}\tilde y^2\partial_{\tilde x}^2\tilde W(\tilde x, \tilde y, t)\,.
  \end{split}
\end{equation}
Apart from the enhanced nonlinear coupling parameter $gs^4$, the squeezing also enhances the dephasing rate to $\gamma_\phi s^4$.
Since both rates scale identically with the squeezing parameter $s$,
we expect the negativity to remain constant as we increase squeezing.
In other words, the effects of dephasing are not exacerbated by increased squeezing.
Also in this case, we can solve Eq.~(\ref{eq:large-squeezing-dephasing-eom}) via a Fourier-transform solution, which is obtained by amending Eq.~(\ref{eq:unitary-fourier-solution}) with the dephasing terms
\begin{equation}
  \label{eq:dephasing-fourier-solution}
  \tilde W(\tilde x,\tilde y,t) = {1\over\sqrt{2\pi}}\int_{-\infty}^\infty dk \, \tilde h_{\tilde y}(k) \,
  \begin{aligned}[t]
     &e^{ik\tilde x-2kgts^4\tilde y^3 - gts^4k^3\tilde y/8} \\
     &\times e^{-\gamma_\phi s^4{\tilde y}^2t/2}
  \end{aligned}
\end{equation}
where $\tilde h_{\tilde y}(k)$ is still defined by Eq.~(\ref{eq:initial-state-fourier-transform}).
We again verify the validity of this approximation by using Eq.~(\ref{eq:dephasing-fourier-solution}) to compute the negativity and compare this to the results from the numerical solution of the full master equation. This is shown in Fig.~\ref{fig:dephasing}d.

\begin{figure}[h]  \includegraphics[width=8.6cm]{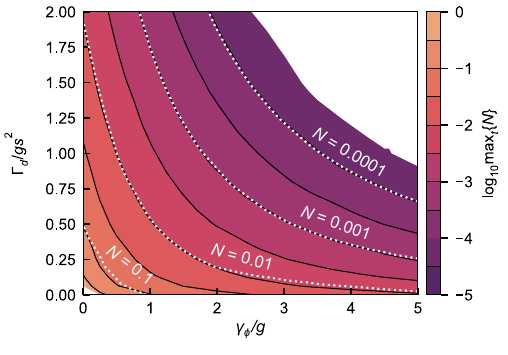}
  \caption{Maximum achieved negativity for systems experiencing both damping and dephasing.
    Colors and black lines constitute a contour plot showing the maximum negativity $\max_tN$ for an initial squeezed state ($s=6.0$) evolved with damping rate, $\Gamma_{\mathrm{d}}$, and dephasing rate, $\gamma_\phi$.
    The white dotted lines show the selected corresponding contours extracted from the solution in Eq.~(\ref{eq:29}) corresponding to the limit of large squeezing, $s\to\infty$. The contour values have been overlayed.
  \label{fig:negativity-contour-plot}}
\end{figure}

We have so far considered two decoherence effects in isolation, but they may be combined by summing the decoherence terms from Eqs.~(\ref{eq:damping-eom}) and (\ref{eq:dephasing-eom}). The transformation to the $\tilde x,\tilde y$-frame and discarding of insignificant terms yields an approximate equation of motion for $\tilde W(\tilde x, \tilde y, t)$ which inherits the three characteristic time scales seen in Eqs.~(\ref{eq:large-squeezing-eom}), (\ref{eq:large-squeezing-damping-eom}), and (\ref{eq:large-squeezing-dephasing-eom}). We now solve the full equation numerically with both mechanisms for varying damping and dephasing rates and compute the resulting maximum negativity. The maximum negativity is shown in Fig.~\ref{fig:negativity-contour-plot}. The corresponding Fourier-transform solution is written
\begin{equation}
  \label{eq:29}
  \tilde W(\tilde x,\tilde y,t) = {1\over\sqrt{2\pi}}\int_{-\infty}^\infty dk \, 
  \begin{aligned}[t]
     &\tilde h_{\tilde y}(k) \,e^{ik\tilde x-2kgts^4\tilde y^3 - gts^4k^3\tilde y/8} \\
     &\times e^{-\Gamma_{\mathrm{d}} s^2t/4-\gamma_\phi s^4{\tilde y}^2t/2}\,.
  \end{aligned}
\end{equation}
Using this, we compute the maximum negativity for both decoherence effects combined and overlay the contours on top of those resulting from the solution of the master equation in Fig.~\ref{fig:negativity-contour-plot}.

\section{Conclusion}\label{sec:conclusions}
In this study, we have investigated the generation of negativity in the Wigner function of a Duffing oscillator, which is initially prepared in a squeezed state. 
Our results demonstrate that the generation rate of negativity is significantly enhanced through squeezing of the initial state. 
We have further explored the impact of two common sources of decoherence, namely energy damping and dephasing, and found that initial squeezing also increases the decoherence rates.
Remarkably, the enhancement of the energy damping rate scales quadratically with the squeezing, which leads to an effective reduction in the thermal decoherence rate compared to the quartic scaling of the nonlinearity. 
Meanwhile, the dephasing rate scales with the quartic power of squeezing, which means that the reduction effect is not present.
A related technique to analyze a similar problem can be found in Ref.~\cite{Riera-Campeny_WignerAnalysisParticle_2023}.

Our findings provide a promising experimental approach for utilizing squeezing as a resource to boost the generation of non-classicality in macroscopic systems.
To identify the most suitable platform for our protocol, Table~\ref{tab:exp-overview} lists recently studied
mechanical systems with their corresponding measured anharmonicity and decoherence rates.
\begin{table*}[!ht]
  \centering
  \begin{tabular}{l@{\hskip 5mm} @{\hskip 5mm}r@{\hskip 5mm} @{\hskip 5mm}r@{\hskip 5mm} H H @{\hskip 5mm}r@{\hskip 5mm} @{\hskip 5mm}r@{\hskip 5mm} @{\hskip 5mm}r@{\hskip 5mm} H @{\hskip 1mm}r@{\hskip 5mm}}
    \hline\hline
    System & ${m~\mathrm{(kg)}}$ & $\Omega_m/(2\pi)\,\mathrm{(Hz)}$ & $T$ & $\gamma_m/(2\pi)\,\mathrm{(Hz)}$ & ${\beta\,\mathrm{(N/m^3)}}$  & ${g/(2\pi)\,\mathrm{(Hz)}}$ & $ {\Gamma_{\mathrm{d}}/(2\pi)\,\mathrm{(Hz)}}$ & \textbf{squeezing} & ${s^2~\mathrm{(dB)}}$ \\ \hline
    C nanotube \cite{eichler_nonlinear_2011,Samanta_NonlinearNanomechanicalResonators_2023}  & $8\times10^{-21}$ & $2.5\times10^{8}$ & $5$ & $1\times10^{4}$ & $6\times10^{12}$ & $1.5$ & $4.2\times10^{6}$ & $1.7\times10^{3}$ & $64$ \\
    SiO nanoparticle \cite{cuairan_precision_2022, tebbenjohanns_quantum_2021} & $3\times10^{-18}$ & $1.3\times10^{5}$ & $3\times10^{2}$ & ~ & $1.2\times10^{7}$& $7.9\times10^{-5}$ & $1\times10^{3}$ & $3.6\times10^{3}$ & $70$ \\
    SiC beam \cite{kozinsky_tuning_2006} & $7.0\times10^{-16}$ & $9.0\times10^{6}$& $4$ & $2.2\times10^{2}$ & $6.3\times10^{14}$ & $1.6\times10^{-5}$ & $2.0\times10^{6}$ & $3.6\times10^{5}$ & $110$ \\
    Si nanobeam \cite{huang_generating_2016}  & $2.0\times10^{-13}$ & $1.6\times10^{6}$ & $6$ & $8.0\times10^{2}$ & $1\times10^{20}$ & $9.8\times10^{-4}$ & $6.3\times10^{7}$ & $2.5\times10^{5}$ & $110$ \\
    AlN beam \cite{aldridge_noise-enabled_2005}  & $3\times10^{-16}$ & $9.3\times10^{7}$ & $4.3$ & $1.3\times10^{4}$ & $3.6\times10^{14}$ & $4.6\times10^{-7}$ & $1.3\times10^{7}$ & $5.2\times10^{6}$ & $130$ \\
    SiN membrane \cite{catalini_modeling_2021-1,seis_ground_2022}  & $1.5\times10^{-11}$ & $1.5\times10^{6}$ & $0.03$ & $1.0\times10^{-3}$ & $1.5\times10^{13}$ &$3.0\times10^{-14}$ & $4.2\times10^{-1}$ & $3.8\times10^{-6}$ & $130$ \\
    Si plate \cite{huang_frequency_2019} & $7.0\times10^{-10}$ & $7.3\times10^{4}$ & $4$ & $8.0\times10^{-2}$ & $4.2\times10^{19}$ & $1.6\times10^{-8}$ & $9.2\times10^{4}$ & $2.4\times10^{6}$ & $130$ \\
    2D materials drum \cite{davidovikj_nonlinear_2017} & $6\times10^{-17}$ &$1.5\times10^{7}$ & $300$ & $1.5\times10^{5}$ & $1.2\times10^{15}$ & $1.5\times10^{-3}$ & $6.3\times10^{10}$ & $6.5\times10^{6}$ & $140$ \\
    SiNx/Gr membrane \cite{singh_giant_2020} & $6.0\times10^{-16}$ & $2.0\times10^{6}$ & $300$ & $2.5\times10^{4}$ & $6\times10^{13}$ & $4.2\times10^{-5}$ & $7.8\times10^{10}$ & $4.3\times10^{7}$ & $150$ \\
    SiN/Nb beam \cite{hocke_determination_2014} & $7.0\times10^{-15}$ & $1.5\times10^{6}$ & $0.4$ & $2.0\times10^{2}$ & $2.1\times10^{11}$ & $1.9\times10^{-9}$ & $1.1\times10^{6}$ & $2.4\times10^{7}$ & $150$ \\
    GaAs nanowire \cite{braakman_nonlinear_2014}  & $5\times10^{-16}$ & $1.33\times10^{6}$ & $4.2$ & $3\times10^{1}$ & $5\times10^{7}$ & $1.1\times10^{-10}$ & $2\times10^{6}$ & $1.3\times10^{8}$& $160$ \\
    Si NEMS \cite{sansa_frequency_2016} & $8.0\times10^{-16}$ & $4.5\times10^{7}$ & $300$ & $7.5\times10^{3}$ & $6.4\times10^{13}$ & $5.0\times10^{-8}$ & $1.0\times10^{9}$ & $1.5\times10^{8}$& $160$ \\
    SiN/Al beam \cite{maillet_measuring_2018}  & $1.0\times10^{-14}$ & $9.0\times10^{5}$ & $4.2$ & $1.5$ & $5\times10^{8}$ & $6.2\times10^{-12}$ & $1.5\times10^{5}$ & $1.5\times10^{8}$ & $160$ \\
    Si nanowire \cite{molina_high_2021} & $7\times10^{-17}$ & $1.9\times10^{6}$ & $300$ & $3.8\times10^{2}$ & $1.4\times10^{7}$ & $7.9\times10^{-10}$ & $1.3\times10^{9}$ & $1.3\times10^{9}$ & $180$ \\
    NEMS \cite{matheny_nonlinear_2013}  & $1.0\times10^{-14}$ & $1.0\times10^{7}$ & $300$ & $1.0\times10^{4}$ & $1\times10^{14}$ & $1.0\times10^{-8}$ & $6.3\times10^{9}$ & $7.9\times10^{8}$ & $180$ \\
    InP membrane \cite{antoni_nonlinear_2012} & $1.0\times10^{-13}$ &$8.0\times10^{5}$ & $300$ & $1.6\times10^{2}$ & $2\times10^{13}$ & $3.1\times10^{-9}$ & $1.3\times10^{9}$ &$6.3\times10^{8}$ & $180$ \\
    Si nanowire \cite{nichol_controlling_2009}  & $7\times10^{-17}$ & $2\times10^{5}$ & $300$ & $6.7\times10^{1}$ & $7\times10^{4}$ & $3.6\times10^{-10}$ & $2.1\times10^{9}$ & $2.4\times10^{9}$ & $190$ \\
    SiN/Au cantilever \cite{villanueva_nonlinearity_2013} & $9.0\times10^{-14}$ & $6.84\times10^{6}$ & $300$ & $2.5\times10^{3}$ & $7.2\times10^{12}$ & $1.9\times10^{-11}$ & $2.3\times10^{9}$ & $1.1\times10^{10}$ & $200$ \\
    GaAs beam \cite{zhang_thermal_2021} & $2.0\times10^{-11}$ & $6.1\times10^{5}$ & $300$ & $1.5\times10^{2}$ & $8\times10^{13}$ & $5.4\times10^{-13}$ & $1.35\times10^{9}$ & $5.4\times10^{10}$ & $210$ \\
    Si wheel \cite{bawaj_probing_2015} & $2.0\times10^{-8}$ & $1.4\times10^{5}$ & $4.3$ & $1.4\times10^{-1}$ & $8\times10^{11}$ & $1.0\times10^{-19}$ & $9.0\times10^{4}$ & $9.4\times10^{11}$ & $240$\\
    Double paddle \cite{bawaj_probing_2015} & $3.3\times10^{-5}$ & $5.6\times10^{3}$ & $293$ & $6.0\times10^{-2}$ & $9.9\times10^{10}$ & $2.9\times10^{-24}$ & $6.6\times10^{7}$ & $4.8\times10^{15}$ & $310$ \\
    \hline\hline
  \end{tabular}
  \caption{Overview of measured anharmonicity in various mechanical resonator systems.
    Comparison of the mass, $m$, mechanical frequency, $\Omega_m/(2\pi)$, Duffing parameter, $\beta$, derived nonlinear coupling parameter, $g$, and damping rate, $\Gamma_{\mathrm{d}}$, of reported mechanical resonator systems sorted by the squeezing in decibels, $s^2$, required for parity between damping and nonlinear parameter, $g s^4 = \Gamma_{\mathrm{d}} s^2$.
  }
  \label{tab:exp-overview}
\end{table*}
From the mass $m$, the resonance frequency $\Omega_m$ and the Duffing constant $\beta$ we calculate the nonlinear rate $g=3\hbar\beta/\left(8m^2\Omega_m^2\right)$.
For all the systems under consideration, the rate $g$ is significantly smaller than the resonance frequency and the decoherence rate $\Gamma_{\mathrm{d}}$, corroborating the assumptions we made throughout in our protocol.

To benchmark the different systems, we evaluate the required amount of squeezing such that the enhanced nonlinear coupling parameter equals the enhanced decoherence rate, i.e., $g s^4 = \Gamma_{\mathrm{d}} s^2$. In this situation, the amount of negativity produced is about $5\times10^{-2}$, according to Fig.~\ref{fig:negativity-contour-plot} at $\gamma_\phi=0$.
This value is similar to the typical error reported in the reconstruction of Wigner functions for bulk acoustic wave resonators~\cite{chu_creation_2018}, indicating the resources required to observe generated negativity.
Table~\ref{tab:exp-overview} also lists the required squeezing in decibels ($10\log_{10}(s^2)$) to achieve this for the various mechanical systems therein.
For reference, to the best of our knowledge, the largest reported squeezing for a mechanical resonator is $s^2\approx6$ \cite{delaney_measurement_2019}.

Of the systems listed, we identify clamped carbon nanotubes and levitated nanoparticles as the two most promising platforms for implementing our protocol.
Clamped carbon nanotubes exhibit a large nonlinearity at the levels of a few tens of mechanical quanta~\cite{Samanta_NonlinearNanomechanicalResonators_2023}, which makes them an attractive option. However, they also suffer from large decoherence rates, necessitating a high level of squeezing ($s^2$ around 60~dB). Whether carbon nanotubes allow for such high levels of squeezing remains an open question.
In contrast, levitated nanoparticles can be isolated effectively from the environment, but the nonlinearity generated by the trapping potential is usually very weak. Therefore, while similar levels of squeezing are required as for carbon nanotubes, the absence of physical clamping allows for the possibility of achieving such high levels of squeezing. Moreover, the high degree of control over the potential energy permits the generation of squeezed states based on parametric modulation or inverted harmonic dynamics~\cite{romero-isart_coherent_2017,neumeier_fast_2022}.
Finally, we expect further reduced decoherence for levitated particles trapped in a dark potential in a cryogenic environment~\cite{romero-isart_quantum_2011}.

This work was supported by the European Research Council project PHOQS (grant no. 101002179), the Novo Nordisk Foundation (grant no. NNF20OC0061866), the Danish National Research Foundation (Center of Excellence ``Hy-Q''), as well as the Independent Research Fund Denmark (grant no. 1026-00345B).

\bibliographystyle{apsrev4-1}
\bibliography{references}

\let\addcontentsline\oldaddcontentsline


\clearpage
\onecolumngrid
\setcounter{equation}{0}
\setcounter{figure}{0}
\setcounter{table}{0}
\setcounter{page}{1}

\renewcommand{\thesection}{\arabic{section}}
\renewcommand{\thesubsection}{\thesection.\arabic{subsection}}
\renewcommand{\thesubsubsection}{\thesubsection.\arabic{subsubsection}}

\phantomsection
\addcontentsline{toc}{title}{Supplementary}
\setcounter{section}{0}

\makeatletter
\renewcommand{\thepage}{S\arabic{page}}
\thispagestyle{plain}
\pagestyle{plain}

\begin{bibunit}[apsrev4-2]
\renewcommand{\theequation}{S\arabic{equation}}
\renewcommand{\thefigure}{S\arabic{figure}}
\renewcommand{\bibnumfmt}[1]{[S#1]}
\renewcommand{\citenumfont}[1]{S#1}

\renewenvironment{widetext}{}{}

\begin{center}
\textbf{\large Supplementary information: Quadrature squeezing enhances Wigner negativity in a mechanical Duffing oscillator}
\end{center}

\vskip2cm
\makebox[\linewidth][c]{\mbox{\begin{minipage}{0.7\linewidth}
      \tableofcontents
    \end{minipage}}}
\vskip2cm

\section{Derivation of the interaction-picture Hamiltonian\label{sec:rwa-hamiltonian}}
We model a nonlinear mechanical oscillator with the Hamiltonian, as given in the main text,
\begin{equation}
\hat{H}=\frac{\hat{p}^{2}}{2m}+\frac{1}{2}m\Omega_m^{2}\hat{q}^{2}+\frac{\beta}{4}\hat{q}^{4}\,,\label{eq:-36}
\end{equation}
where $\hat q$ and $\hat p$ describe the position and momentum of the mechanical degree of freedom, $m$ describes the mass, and $\Omega_m$ describes the resonance frequency. The Duffing parameter, $\beta$, here has dimensions of $(\text{force})\times(\text{length})^{-3}$ and describes the change in stiffness with displacement. For context, we note that the derived Heisenberg operator equations of motion are
\begin{equation}
\dot{\hat{q}}=\frac{\hat{p}}{m}\label{eq:-68}
\end{equation}
and
\begin{equation}
\dot{\hat{p}}=-m\Omega_m^{2}\hat{q}-m\gamma\dot{\hat{q}}-\beta\hat{q}^{3},\label{eq:-69}
\end{equation}
where we have additionally introduced a linear damping term with damping rate $\gamma$.
Introducing the ladder operators $\hat{a}$ and $\hat{a}^{\dagger}$ by a transformation of the position and momentum operators,
\begin{equation}
  \label{eq:-180}
  \hat{a}=\sqrt{\frac{1}{2\hbar m\Omega_m}}\left(m\Omega_m\hat{q}+i\hat{p}\right)
\end{equation}
and
\begin{equation}
  \hat{a}^{\dagger}=\sqrt{\frac{1}{2\hbar m\Omega_m}}\left(m\Omega_m\hat{q}-i\hat{p}\right),
\end{equation}
we can rewrite the Hamiltonian as
\begin{equation}
  \label{eq:20}
  \hat H = \hbar\Omega_m\left(\hat{a}^\dagger a + \frac12\right) + \frac{\hbar^2\beta}{16m^2\Omega_m^2}\left(\hat{a}+\hat{a}^\dagger\right)^4\,.
\end{equation}

Since the nonlinear dynamics are generally weak (defined below), we transform the Hamiltonian to the interaction picture and apply the rotating wave approximation. To change to a frame rotating with frequency $\omega_0$, we make the substitutions~\cite{Sakurai_ModernQuantumMechanics_2011a}
\begin{align}
\hat{a} & \to\hat{a}_{I}=e^{i\hat{H}_{0}t}\hat{a}e^{-i\hat{H}_{0}t}=\hat{a}e^{-i\omega_{0}t}\,,\\
\hat{a}^{\dagger} & \to\hat{a}_{I}^{\dagger}=e^{i\hat{H}_{0}t}\hat{a}^{\dagger}e^{-i\hat{H}_{0}t}=\hat{a}^{\dagger}e^{i\omega_{0}t}\,.
\end{align}
while subtracting the base Hamiltonian, $H_0 = \hbar\omega_0(\hat{a}^\dagger a+1/2)$, to obtain the following interaction-picture Hamiltonian, $\hat H_I$:
\begin{align}
\begin{aligned}[b] \hat{H}_{I} & = \frac{3\hbar^{2}\beta}{8m^{2}\omega_{0}^{2}}\left(\hat{a}^{\dagger}\hat{a}^{\dagger}\hat{a}\hat{a}+2\hat{a}^{\dagger}\hat{a}+\frac{1}{2}\right)+\frac{\hbar^{2}\beta}{8m^{2}\omega_{0}^{2}}\left[\left(2\hat{a}^{\dagger}\hat{a}^{3}+3\hat{a}^{2}\right)e^{-2i\omega_{0}t}+\mathrm{h.c.}\right]\\
 &\qquad +\frac{\hbar^{2}\beta}{16m^{2}\omega_{0}^{2}}\left(\hat{a}^{4}e^{-4i\omega_{0}t}+\mathrm{h.c.}\right)+\hbar(\Omega_m-\omega_{0})\left(\hat{a}^{\dagger}\hat{a}\right).
\end{aligned}
\label{eq:-13}
\end{align}
The symbol ``h.c.'' denotes the Hermitian conjugate of the other terms within its innermost containing parentheses.

For a weak nonlinearity, we have $\Omega_m \gg g$, i.e. the investigated dynamics are slow compared to the system base oscillator frequency, $\Omega_m$.
From this, we see that by choosing $\omega_0$ as the real root of the polynomial $4m^2\omega_0^2(\Omega_m-\omega_0)+3\hbar^2\beta$ and omitting constant terms, we arrive at the sought after Hamiltonian used in the main text,
\begin{equation}
  \label{eq:17}
  \hat{H}=\hbar g\hat{a}^{\dagger}\hat{a}^{\dagger}\hat{a}\hat{a} = \hbar g(\hat n^2 - \hat n)\,,
\end{equation}
with the nonlinear coupling parameter identified as
\begin{equation}
  \label{eq:21}
  g=\frac{3\hbar\beta}{8m^{2}\Omega_m^{2}}.
\end{equation}
This final equation, along with the condition that $\Omega_m\gg g$, defines whether the nonlinearity is weak in relation to $\beta$, as used above.

\section{Derivation of Wigner function equation of motion}
We consider in the main text, a system described by the Hamiltonian (\ref{eq:17}) subject to varying selections of the decoherence processes of linear damping and dephasing. The most general master equation, which describes the evolution of the density matrix, $\hat\rho$, under the system dynamics described by the Hamiltonian (\ref{eq:17}) and both linear damping and dephasing is given by~\cite{Gardiner_QuantumNoise_2004,Wilson_MeasurementbasedControlMechanical_2015}
\begin{equation}
  \label{eq:1}
\dot{\hat{\rho}}(t)  = -ig\left[\hat n^2 - \hat n,\hat{\rho}(t)\right]
 +\gamma(\bar{n}+1)\mathcal{D}\left[\hat{a}\right]\hat{\rho}(t)+\gamma\bar{n}\mathcal{D}\left[\hat{a}^{\dagger}\right]\hat{\rho}(t)+\gamma_{\phi}\mathcal{D}\left[\hat{n}\right]\hat{\rho}(t).
\end{equation}
where
\begin{align}
  \label{eq:16}
  \mathcal{D}\left[\hat{a}\right]\hat{\rho}(t) &= \hat{a}\hat{\rho}\hat{a}^{\dagger}-\tfrac{1}{2}\hat{a}^{\dagger}\hat{a}\hat{\rho}-\tfrac{1}{2}\hat{\rho}\hat{a}^{\dagger}\hat{a}\,,
  \\\label{eq:18}
  \mathcal{D}\left[\hat{a}^{\dagger}\right]\hat{\rho}(t) &= \hat{a}^\dagger\hat{\rho}\hat{a}-\tfrac{1}{2}\hat{a}\hat{a}^\dagger\hat{\rho}-\tfrac{1}{2}\hat{\rho}\hat{a}\hat{a}^\dagger\,,
  \\\label{eq:19}
  \mathcal{D}\left[\hat{n}\right]\hat{\rho}(t) &= \hat{n}\hat{\rho}\hat{n}^{\dagger}-\tfrac{1}{2}\hat{n}^{\dagger}\hat{n}\hat{\rho}-\tfrac{1}{2}\hat{\rho}\hat{n}^{\dagger}\hat{n}\,.
\end{align}
Equation~(\ref{eq:1}) describes a nonlinear oscillator which experiences linear damping through coupling to a thermal bath of mean occupancy $\bar{n}_\mathrm{th}$ with coupling rate $\gamma$ and dephasing at a rate $\gamma_\phi$. The parameter $g$ is introduced in Supplementary Section~\ref{sec:rwa-hamiltonian} and describes the strength of the nonlinearity.

\subsection{General procedure}
A procedure used to derive the Fokker-Planck-like equation of motion for the Wigner function from the corresponding master equation is given in Ref.~\cite{Walls_QuantumOptics_2008}: The characteristic function corresponding to the density matrix $\hat\rho$ is defined as
\begin{equation}
  \label{eq:4}
  \chi(\lambda,\lambda^{*},t) = \operatorname{Tr}\left[\hat{\rho}(t)\hat{D}(\lambda)\right]\,,
\end{equation}
and relates to the Wigner function by the complex Fourier transformation,
\begin{equation}
  \label{eq:3}
  W(\alpha,\alpha^{*},t)=\frac{1}{\pi^{2}}\int_{\mathbb{C}} d\lambda\,d\lambda^{*}\,e^{\lambda^{*}\alpha-\alpha^{*}\lambda}\chi(\lambda,\lambda^{*},t).
\end{equation}
where $\hat D(\lambda)$ denotes the displacement operator, $\hat D(\lambda) = e^{\lambda \hat a^\dagger - \lambda^* \hat a}$.
Starting from Eq.~(\ref{eq:1}), we derive an equation of motion for $W(\alpha,\alpha^{*},t)$ as
\begin{equation}
  \label{eq:-128}
  \partial_tW(\alpha,\alpha^{*},t)
  = \frac{1}{\pi^{2}}\int_{\mathbb{C}} d\lambda\,d\lambda^{*}\,\partial_{t}\chi(\lambda,\lambda^{*},t)
  = \frac{1}{\pi^{2}}\int_{\mathbb{C}} d\lambda\,d\lambda^{*}\,\operatorname{Tr}\left[\dot{\hat{\rho}}(t)\hat{D}(\lambda)\right]\,.
\end{equation}
in which the equation of motion for $\partial_{t}\chi(\lambda,\lambda^{*},t)$ is derived as
\begin{equation}
  \label{eq:8}
  \partial_{t}\chi(\lambda,\lambda^{*},t)
  = \operatorname{Tr}\left[\dot{\hat{\rho}}(t)\hat{D}(\lambda)\right]\,.
\end{equation}
After substituting $\dot{\hat\rho}(t)$ using the desired master equation, e.g. Eq.~(\ref{eq:1}), into Eq.~(\ref{eq:-128}), the following identities are used to express the right-hand side in terms of spatial derivatives of $\chi(\lambda,\lambda^*,t)$:
\begin{align}
\hat{a}\hat{D}(\lambda) & =\left(-\partial_{\lambda^{*}}+\frac{\lambda}{2}\right)\hat{D}(\lambda),\label{eq:-6}\\
\hat{a}^{\dagger}\hat{D}(\lambda) & =\left(\partial_{\lambda}+\frac{\lambda^{*}}{2}\right)\hat{D}(\lambda),\label{eq:27}\\
\hat{D}(\lambda)\hat{a}^{\dagger} & =\left(\partial_{\lambda}-\frac{\lambda^{*}}{2}\right)\hat{D}(\lambda),\label{eq:-7}\\
\hat{D}(\lambda)\hat{a} & =-\left(\partial_{\lambda^{*}}+\frac{\lambda}{2}\right)\hat{D}(\lambda)\,.\label{eq:-8}
\end{align}
Finally, a partial differential equation for $W(\alpha,\alpha^*,t)$ is obtained by rewriting each term on the resulting right-hand side of Eq.~(\ref{eq:-128}) according to the prescription
\begin{equation}
  \label{eq:-79-1}
  \begin{aligned}[b]
    \frac{1}{\pi^{2}}\int_{\mathbb{C}} d\lambda\,d\lambda^{*}\, e^{\alpha\lambda^{*}-\alpha^{*}\lambda}\lambda^{m}\left(\lambda^{*}\right)^{n}\partial_{\lambda}^{p}\partial_{\lambda^{*}}^{q}\chi(\lambda,\lambda^{*},t)
    & =\frac{1}{\pi^{2}}(-1)^{m}\partial_{\alpha^{*}}^{m}\partial_{\alpha}^{n}\int_{\mathbb{C}} d\lambda\,d\lambda^{*}\,e^{\alpha\lambda^{*}-\alpha^{*}\lambda}\partial_{\lambda}^{p}\partial_{\lambda^{*}}^{q}\chi(\lambda,\lambda^{*},t) \\
    & =\frac{1}{\pi^{2}}(-1)^{m+p+q}\partial_{\alpha^{*}}^{m}\partial_{\alpha}^{n}\left[\int_{\mathbb{C}} d\lambda\,d\lambda^{*}\,\partial_{\lambda}^{p}\partial_{\lambda^{*}}^{q}e^{\alpha\lambda^{*}-\alpha^{*}\lambda}\chi(\lambda,\lambda^{*},t) \right]\\
    & =(-1)^{m+q}\partial_{\alpha^{*}}^{m}\partial_{\alpha}^{n}\left[\left(\alpha^{*}\right)^{q}\alpha^{p}W(\alpha,\alpha^{*},t)\right].
  \end{aligned}
\end{equation}

Since the procedure is linear with respect to the terms of the master equation, we may consider terms separately and combine them later. Here, we recast the obtained equations in terms of the quadrature coordinates, $x$ and $y$, using the coordinate transformation given by
\begin{equation}
  \label{eq:7}
  \alpha = x + i y\qquad\text{and}\qquad \alpha^* = x - iy\,.
\end{equation}

\subsection{Unitary evolution}
We consider first the unitary evolution given by
\begin{equation}
  \label{eq:13}
\dot{\hat{\rho}}=-ig\left[\hat n^2 - \hat n,\hat{\rho}\right].
\end{equation}
Using Eqs.~(\ref{eq:8}--\ref{eq:-8}), we find
\begin{equation}
\label{eq:-245}
\begin{aligned}
  \partial_{t}\chi(\lambda,\lambda^{*},t)
  = 2ig\left(-\lambda\partial_{\lambda^{*}}\partial_{\lambda}^{2}+\lambda^{*}\partial_{\lambda^{*}}^{2}\partial_{\lambda}\right)\chi(\lambda,\lambda^{*},t)
  -\frac{ig}{2}\left(-\lambda^{2}\lambda^{*}\partial_{\lambda}+\lambda\left(\lambda^{*}\right)^{2}\partial_{\lambda^{*}}\right)\chi(\lambda,\lambda^{*},t)\,,
\end{aligned}
\end{equation}
to which applying Eq.~(\ref{eq:-79-1}) yields
\begin{equation}
  \label{eq:-246}
  \partial_{t}W(\alpha,\alpha^{*})
  =2ig\left(\alpha^{2}\alpha^{*}\partial_{\alpha}-\alpha\left(\alpha^{*}\right)^{2}\partial_{\alpha^{*}}\right)W(\alpha,\alpha^{*})
 -\frac{ig}{2}\left(\alpha\partial_{\alpha^{*}}\partial_{\alpha}^{2}-\alpha^{*}\partial_{\alpha}\partial_{\alpha^{*}}^{2}\right)W(\alpha,\alpha^{*})\,.
\end{equation}
Expressed in the quadrature coordinates given by Eq.~(\ref{eq:7}), the equation of motion for $W$, subject only to unitary evolution, becomes
\begin{equation}
  \label{eq:12}
\partial_{t}W(x,y)  =2g\left(x^{2}+y^{2}\right)\left(-y\partial_{x}+x\partial_{y}\right)W(x,y)-\frac{g}{8}\left(-y\partial_{x}+x\partial_{y}\right)\left(\partial_{x}^{2}+\partial_{y}^{2}\right)W(x,y)\,.
\end{equation}

\subsection{Damping}
Secondly, we consider the case of damping through coupling to a thermal bath of temperature $\bar{n}_\mathrm{th}$ and with coupling strength $\gamma$ as introduced in the main text. The master equation is obtained by keeping only the second and third terms on the right-hand side of Eq.~(\ref{eq:1}).
\begin{equation}
  \label{eq:11}
\dot{\hat{\rho}}=\gamma(\bar{n}_\mathrm{th}+1)\left(\hat{a}\hat{\rho}\hat{a}^{\dagger}-\frac{1}{2}\hat{a}^{\dagger}\hat{a}\hat{\rho}-\frac{1}{2}\hat{\rho}\hat{a}^{\dagger}\hat{a}\right)+\gamma\bar{n}_\mathrm{th}\left(\hat{a}^{\dagger}\rho\hat{a}-\frac{1}{2}\hat{a}\hat{a}^{\dagger}\hat{\rho}-\frac{1}{2}\hat{\rho}\hat{a}\hat{a}^{\dagger}\right).
\end{equation}
Again, using Eqs.~(\ref{eq:8}--\ref{eq:-8}), we find
\begin{equation}
\partial_{t}\chi(\lambda,\lambda^{*},t) =-\gamma\left(\bar{n}+\frac{1}{2}\right)\lambda\lambda^{*}\chi(\lambda,\lambda^{*},t) -\gamma\lambda\partial_{\lambda}\chi(\lambda,\lambda^{*},t) -\gamma\lambda^{*}\partial_{\lambda^{*}}\chi(\lambda,\lambda^{*},t) \label{eq:-251}
\end{equation}
to which applying Eq.~(\ref{eq:-79-1}) yields
\begin{equation}
\partial_{t}W(\alpha,\alpha^{*},t) = \Gamma_{\mathrm{d}}\partial_{\alpha}\partial_{\alpha^{*}}W(\alpha,\alpha^{*},t)+\frac{\gamma}{2}\partial_{\alpha}\left(\alpha W(\alpha,\alpha^{*},t)\right)+\frac{\gamma}{2}\partial_{\alpha^{*}}\left(\alpha^{*}W(\alpha,\alpha^{*},t)\right)\label{eq:-89}
\end{equation}
where we have used $\Gamma_{\mathrm{d}}=\gamma(\bar{n}_\mathrm{th} + 1/2)$. In the quadrature coordinates, $x$ and $y$, Eq.~(\ref{eq:-89}) takes the form
\begin{equation}
  \label{eq:-92}
\partial_{t}W(x,y,t)= \frac{\Gamma_{\mathrm{d}}}{4}\left(\partial_{x}^{2}+\partial_{y}^{2}\right)W(x,y,t)+\frac{\gamma}{2}\partial_{x}\left(xW(x,y,t)\right)+\frac{\gamma}{2}\partial_{y}\left(yW(x,y,t)\right)
\end{equation}
For large temperatures, i.e. $\bar{n}_\mathrm{th}\gg1$, only the first term of Eq.~(\ref{eq:-89}) is significant.

\subsection{Dephasing}
As an additional decoherence process, we consider dephasing with a rate $\gamma_\phi$, whose master equation is
\begin{equation}
  \dot{\hat{\rho}}_{\gamma_{\phi}}=\gamma_{\phi}\left(\hat{n}\hat{\rho}\hat{n}-\frac{1}{2}\hat{n}^{2}\hat{\rho}-\frac{1}{2}\hat{\rho}\hat{n}^{2}\right)\,.
\end{equation}
Again, using Eqs.~(\ref{eq:8}--\ref{eq:-8}), we find
\begin{equation}
\partial_{t}\chi(\lambda,\lambda^{*},t) =-\frac{\gamma_{\phi}}{2}\lambda^{2}\partial_{\lambda}^{2}\chi(\lambda,\lambda^{*},t) -\frac{\gamma_{\phi}}{2}\left(\lambda^{*}\right)^{2}\partial_{\lambda^{*}}^{2}\chi(\lambda,\lambda^{*},t) +\gamma_{\phi}\lambda\lambda^{*}\partial_{\lambda}\partial_{\lambda^{*}}\chi(\lambda,\lambda^{*},t) .\label{eq:-250}
\end{equation}
to which applying Eq.~(\ref{eq:-79-1}) yields
\begin{equation}
  \label{eq:-247}
  \partial_{t}W(\alpha,\alpha^{*},t) =-\frac{\gamma_{\phi}}{2}\left(\alpha\partial_{\alpha}-\alpha^{*}\partial_{\alpha^{*}}\right)^{2}W(\alpha,\alpha^{*},t).
\end{equation}
In the quadrature coordinates, dephasing takes the form
\begin{equation}
  \label{eq:6}
\partial_{t}W(x,y,t)=\frac{\gamma_{\phi}}{2}\left(-y\partial_{x}+x\partial_{y}\right)^{2}W(x,y,t).
\end{equation}
We can think of this as a diffusion of the Wigner function in the angular coordinate.

\subsection{Equation of motion for nonlinear oscillator with decoherence effects}
The equation of motion for the Wigner function, $W$, given the master equation (\ref{eq:1}) can be determined by linearly combining the right-hand-side terms of Eqs.~(\ref{eq:12}), (\ref{eq:-92}), and (\ref{eq:6}). This yields (cf. Refs.~\cite{Stobinska_WignerFunctionEvolution_2008,Walls_QuantumOptics_2008}) the equation
\begin{equation}
  \label{eq:2}
  \begin{aligned}[b]
    \partial_{t}W(x,y) &=
    2g\left(x^{2}+y^{2}\right)\left(-y\partial_{x}+x\partial_{y}\right)W(x,y)-\frac{g}{8}\left(-y\partial_{x}+x\partial_{y}\right)\left(\partial_{x}^{2}+\partial_{y}^{2}\right)W(x,y)
\\&\qquad +\frac{\Gamma_{\mathrm{d}}}{4}\left(\partial_{x}^{2}+\partial_{y}^{2}\right)W(x,y,t)+\frac{\gamma}{2}\partial_{x}\left(xW(x,y,t)\right)+\frac{\gamma}{2}\partial_{y}\left(yW(x,y,t)\right)
\\&\qquad +\frac{\gamma_{\phi}}{2}\left(-y\partial_{x}+x\partial_{y}\right)^{2}W(x,y,t)\,,
  \end{aligned}
\end{equation}
where the first line describes the unitary dynamics of a nonlinear oscillator with nonlinearity rate, $g$, the second line describes the dynamics of linear damping, and the third line describes the dynamics of dephasing given a dephasing rate $\gamma_\phi$.

\section{Approximate dynamics for a squeezed initial state}
We consider in this work as the main problem the evolution of a squeezed vacuum state under the dynamics described by Eq.~(\ref{eq:5}). To derive an approximate solution, the equation of motion for the Wigner function, $W$, is transformed to the rescaled quadrature coordinates,
\begin{equation}
  \label{eq:10}
  \tilde x=sx \qquad\text{and}\qquad \tilde y = y/s\,,
\end{equation}
where $s$ is the squeezing parameter for the initial state as described in the main text. This transformation allows us to neglect several terms on the right-hand side, keeping only those which are significant for the initial short-time evolution. Here, we list the exact equations of motion for the Wigner function, $W$, in the scaled coordinates, $\tilde x$ and $\tilde y$, before the terms are neglected as in the main text. Starting from the equation of motion (\ref{eq:5}), and implementing the substitutions given by Eq.~(\ref{eq:10}), as well as the therefrom derived substitutions
\begin{equation}
  \label{eq:14}
  \partial_{\tilde{x}}=\frac{1}{s}\partial_{x}\qquad\text{and}\qquad\partial_{\tilde{y}}=s\partial_{y}\,,
\end{equation}
we obtain
\begin{equation}
  \label{eq:5}
  \begin{aligned}[b]
    \partial_t \tilde W(\tilde x,\tilde y,t)
    &= 2g\Big(-\tilde x^2\tilde y\partial_{\tilde x} - s^4\tilde y^3\partial_{\tilde x} + {1\over s^4}\tilde x^3\partial_{\tilde y} + \tilde x\tilde y^2\partial_{\tilde y}\Big)\tilde W(\tilde x,\tilde y,t) +{g\over8}\Big(-s^4\tilde y\partial_{\tilde x}^3 + {1\over s^4}\tilde x\partial_{\tilde y}^3 + \tilde x\partial_{\tilde y}\partial_{\tilde x}^2 - \tilde y\partial_{\tilde x}\partial_{\tilde y}^2\Big)\tilde W(\tilde x,\tilde y,t) \\
    &\phantom{=} + \frac{\Gamma_{\mathrm{d}}s^2}{4}\partial_{\tilde x}^2\tilde W(\tilde x, \tilde y, t) + \frac{\Gamma_{\mathrm{d}}}{4s^2}\partial_{\tilde y}^2\tilde W(\tilde x, \tilde y, t) + {\gamma\over 2}\partial_{\tilde x}\left(\tilde x\tilde W(\tilde x, \tilde y, t)\right) + {\gamma\over2}\partial_{\tilde y}\left(\tilde y\tilde W(\tilde x, \tilde y, t)\right) \\
    &\phantom{=} + \frac{\gamma_{\phi}}{2}\left(s^{4}\tilde{y}^{2}\partial_{\tilde{x}}^{2}+s^{-4}\tilde{x}^{2}\partial_{\tilde{y}}^{2}-2\tilde{x}\tilde{y}\partial_{\tilde{x}}\partial_{\tilde{y}}-\tilde{x}\partial_{\tilde{x}}-\tilde{y}\partial_{\tilde{y}}\right)\tilde{W}(\tilde{x},\tilde{y},t) \,.
  \end{aligned}
\end{equation}
We consider in this work, the regime of large squeezing, i.e., $s\gg1$. Thus, we retain only terms with the highest power of $s$ in combination with each of the coefficients $g$, $\Gamma_{\mathrm{d}}$, and $\gamma_\phi$ separately. Neglecting other terms, we arrive at the following equation of motion given large $s$:
\begin{equation}
  \label{eq:15}
    \partial_t \tilde W(\tilde x,\tilde y,t)
    = -2gs^4\tilde y^3\partial_{\tilde x}\tilde W(\tilde x,\tilde y,t)
  -{gs^4\over8}\tilde y\partial_{\tilde x}^3\tilde W(\tilde x,\tilde y,t)
  +  \frac{\Gamma_{\mathrm{d}}s^2}{4}\partial_{\tilde x}^2\tilde W(\tilde x, \tilde y, t)
   + {\gamma_\phi s^4\over 2}\tilde y^2\partial_{\tilde x}^2\tilde W(\tilde x, \tilde y, t)\,.
\end{equation}
Since $\gamma=\Gamma_{\mathrm{d}}/(\bar{n}_{\mathrm{th}}+1/2)$, terms prefixed by $\gamma$ have also been neglected with reference to the term prefixed by $\Gamma_{\mathrm{d}}s^2$.

The equations presented in the main text are special cases of Eq.~(\ref{eq:15}) and can be obtained by setting selected coefficients to zero: Set $\gamma_\phi=0$ to obtain the equation of motion for the system which experiences only nonlinear dynamics and damping and set $\Gamma_{\mathrm{d}}=0$ to obtain the equation of motion for the system which experiences only nonlinear dynamics and dephasing. Using Eq.~(\ref{eq:15}) or the derived equations, we can directly inspect the relative scaling between the different parts of system dynamics, namely that the nonlinear unitary dynamics as well as dephasing scale as $s^4$ while damping only scales as $s^2$.

\begin{figure}
  \includegraphics[width=17.8cm]{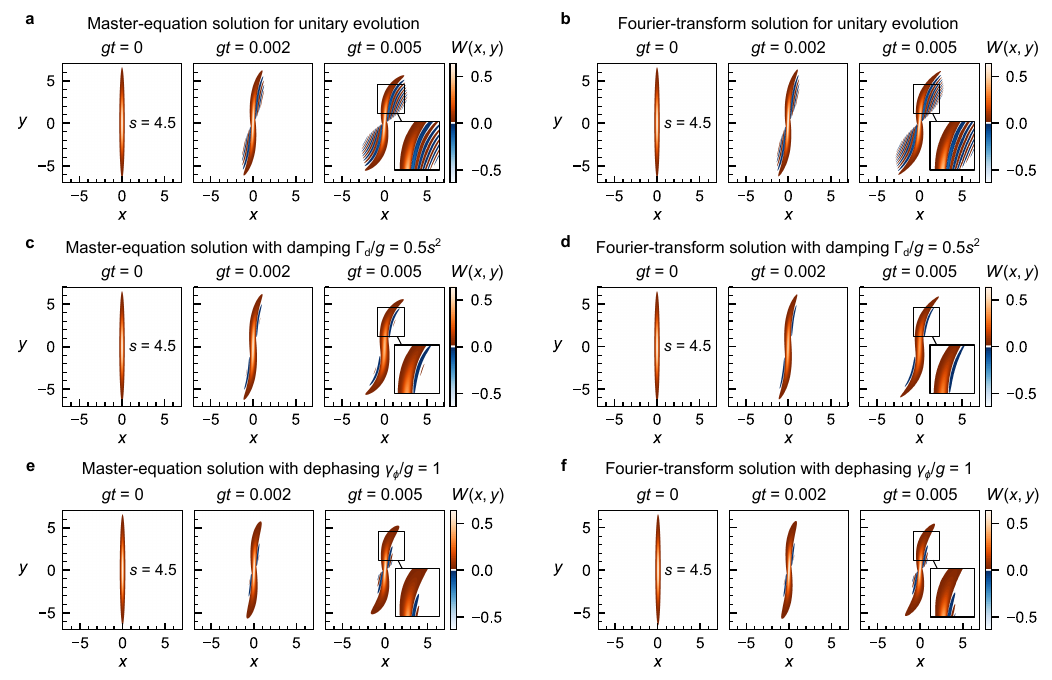}
  \caption{Side-by-side comparison between Wigner functions computed using the full master equation solution and using the approximate Fourier-transform solution.
    \textbf{a}, Unitary evolution, i.e. $\Gamma_{\mathrm{d}}=0$ and $\gamma_\phi=0$, of the Wigner function of a squeezed state with $s=4.5$ as simulated by the full master equation (\ref{eq:1}).
    \textbf{b}, Unitary evolution of the Wigner function with parameters as in \textbf{a} computed using the Fourier transform (\ref{eq:9}).
    \textbf{c}, Evolution of a squeezed state with linear damping with parameters $s=4.5$, $\Gamma_{d}=0.5s^2$, $\bar{n}_{\mathrm{th}}=1000$, and $\gamma_\phi=0$ simulated using the full master equation (\ref{eq:1}).
    \textbf{d}, Evolution of the Wigner function under linear damping with parameters as in \textbf{c} computed using the Fourier transform (\ref{eq:9}).
    \textbf{e}, Evolution of a squeezed state with linear damping with parameters $s=4.5$, $\Gamma_{d}=0$ and $\gamma_\phi=1$ simulated using the full master equation (\ref{eq:1}).
    \textbf{f}, Evolution of the Wigner function under dephasing with parameters as in \textbf{e} computed using the Fourier transform (\ref{eq:9}).
  }
    \label{fig:asymp-comparison}
\end{figure}

As noted in the main text, the solution to Eq.~(\ref{eq:15}) may be expressed as the Fourier transform
\begin{equation}
  \label{eq:9}
  \tilde W(\tilde x,\tilde y,t) = {1\over\sqrt{2\pi}}\int_{-\infty}^\infty dk \,
  \tilde h_{\tilde y}(k) \,e^{ik\tilde x-2kgts^4\tilde y^3 - gts^4k^3\tilde y/8}
  \times e^{-\Gamma_{\mathrm{d}} s^2t/4-\gamma_\phi s^4{\tilde y}^2t/2}\,,
\end{equation}
where $h_{\tilde y}(k)$ is found from the initial state, $\tilde W(\tilde x,\tilde y,0)$, as
$\tilde h_{\tilde y}(k) = (2\pi)^{-1/2} \int_{-\infty}^\infty d\tilde x\, \tilde W(\tilde x,\tilde y,0)e^{-ik\tilde x}$. To illustrate the validity of this approximate solution, Fig.~\ref{fig:asymp-comparison} shows a side-by-side comparison of the Wigner functions computed from the solution of the full master equation (as shown in the main text) and computed from the Fourier-transform expression above.

\section{Decay of negativity in finite time}
In a system subject only to linear damping, i.e., governed by the master equation (\ref{eq:11}) or, equivalently, the Wigner function equation of motion (\ref{eq:-92}), it can be shown that all negativity of the Wigner function will vanish after a finite time. Equation~(\ref{eq:-92}) can be recognized as the Fokker-Planck equation of an Ornstein--Uhlenbeck process in two spatial dimensions \cite{Gardiner_HandbookStochasticProcesses_1985} with diffusion coefficient $\Gamma_s/2$ and drift coefficient ${\gamma}/{2}$. For an initial (unphysical) Wigner function given by the delta function
\begin{equation}
  \label{eq:26}
  W_{\delta}(x,y,0)=\delta(x-x_{0})\delta(y-y_{0})\,,
\end{equation}
Eq.~(\ref{eq:-92}) is solved by a Gaussian \cite{Wang_TheoryBrownianMotion_1945} with expectation values
\begin{align}
  \langle x\rangle_{t} & =\int dx\,dy\,xW_{\delta}(x,y,t)=x_{0}e^{-\gamma t/2}\qquad\text{with \({x_{0}=\langle x\rangle_{t=0}}\;,\)}\label{eq:22}\\
  \langle y\rangle_{t} & =\int dx\,dy\,yW_{\delta}(x,y,t)=y_{0}e^{-\gamma t/2}\qquad\text{with \({y_{0}=\langle y\rangle_{t=0}}\;\),}\label{eq:23}
\end{align}
and (co)variances
\begin{align}
  \langle(\Delta x)^{2}\rangle_{t} & =\int dx\,dy\,\left(x^{2}-x\langle x\rangle_{t}\right)W_{\delta}(x,y,t)=\frac{\Gamma_{\mathrm{d}}}{2\gamma}\left(1-e^{-\gamma t}\right),\label{eq:-195}\\
  \langle(\Delta y)^{2}\rangle_{t} & =\int dx\,dy\,\left(y^{2}-y\langle y\rangle_{t}\right)W_{\delta}(x,y,t)=\frac{\Gamma_{\mathrm{d}}}{2\gamma}\left(1-e^{-\gamma t}\right),\label{eq:-196}\\
  \left\langle \left(x-\langle x\rangle_{t}\right)\left(y-\langle y\rangle_{t}\right)\right\rangle _{t} & =\int dx\,dy\,\left(x-\langle x\rangle_{t}\right)\left(y-\langle y\rangle_{t}\right)W_{\delta}(x,y,t)=0\,.\label{eq:24}
\end{align}
Equations (\ref{eq:26})--(\ref{eq:24}) allows us to express the solution to Eq.~(\ref{eq:-92}) given an arbitrary initial state $W(x,y,0)$ with
\begin{equation}
  W(xe^{-\gamma t/2},ye^{-\gamma t/2},t)=\int dx'\,dy'\,W(x',y',0)\exp\left(\frac{-(x-x')^{2}-(y-y')^{2}}{(\Gamma_{\mathrm{d}}/\gamma)\left(1-e^{-\gamma t}\right)e^{\gamma t}}\right).\label{eq:-189}
\end{equation}

Using the Husimi Q function, defined as
\begin{equation}
  \label{eq:25}
  Q(\alpha,\alpha^{*})=\langle\alpha|\hat{\rho}|\alpha\rangle\,,
\end{equation}
we can establish a finite bound for the time evolved under Eq.~(\ref{eq:-92}) after which the Wigner function is completely non-negative. The Q function is related to the Wigner function through the convolution
\begin{equation}
  Q(\beta,\beta^{*})=\int d\alpha\,d\alpha^{*}\,W(\alpha,\alpha^{*})e^{-2|\alpha-\beta|^{2}}.\label{eq:-188}
\end{equation}

Comparing Eqs.~(\ref{eq:-188}) and (\ref{eq:-189}), we see that defining
\begin{equation}
  \label{eq:-194}
  t_{\mathrm{decay}} =\gamma^{-1}\log\left(1+\frac{\gamma}{2\Gamma_{\mathrm{d}}}\right) =\gamma^{-1}\log\left(1+\frac{1}{2\bar{n}+1}\right)\,,
\end{equation}
we have
\begin{equation}
  W(x e^{-\gamma t_{\mathrm{decay}}/2},y e^{-\gamma t_{\mathrm{decay}}/2},t_{\mathrm{decay}})=Q(x,y,0).
\end{equation}
From its definition (\ref{eq:25}), the Q function is manifestly non-negative for all states. Thus, the Wigner function must also be non-negative at the characteristic time $t_{\mathrm{decay}}$, which is finite for any nonzero $\gamma$. Since the evolution of a non-negative Wigner function under Eq.~(\ref{eq:-92}) can never lead to negativity, the Wigner function remains non-negative at all times greater than $t_{\mathrm{decay}}$.

\section[Numerics]{Numerical evaluation of the Wigner function}
To compute the Wigner negativity as a function of time for the full system, we solve the equivalent master equation (\ref{eq:1}) using QuTiP~\cite{johansson_qutip_2013}. For the unitary case, the time-evolution of an initial state in the number state basis is trivially solvable
\begin{equation}
  \label{eq:28}
  |\Psi(t)\rangle = \sum_n c_ne^{-i(n^2-n)t}|n\rangle\,.
\end{equation}
For the non-unitary evolution, we numerically solve the master equation (\ref{eq:1}) in a truncated Fock basis with the first $n_{\max}$ states. From the states $|\Psi(t)\rangle$ and $\hat\rho(t)$, we use QuTiP to compute the Wigner function using transition probabilities~\cite{Curtright_ConciseTreatiseQuantum_2014,Bartlett_ExactTransitionProbabilities_1949}.  The Wigner function is evaluated on a square grid of $N_x$-by-$N_x$ points,  $(x_i,y_j)$, centered on the origin $(x,y) = (0,0)$, where adjacent points are separated by the grid resolution, $\Delta x$, i.e.,
$x_{i+1}-x_i=y_{j+1}-y_j=\Delta x$. The grid extent is derived as $x_\text{max} = x_{N_x} = - x_1 = y_{N_x} = - y_1 = N_x\Delta x/2$. Thus, the Wigner negativity, $N$, at time, $t$, is computed as
\begin{equation}
    N(t) = \sum_{i=1}^{N_x}\sum_{j=1}^{N_x}\min\{0,W(x_i, y_j,t)\}
\end{equation}
where $W(x_i, y_j,t)$ is the Wigner function evaluated for the relevant state, $\hat\rho(t)$ or $|\Psi(t)\rangle$.

\begin{figure}
  \includegraphics[width=17.8cm]{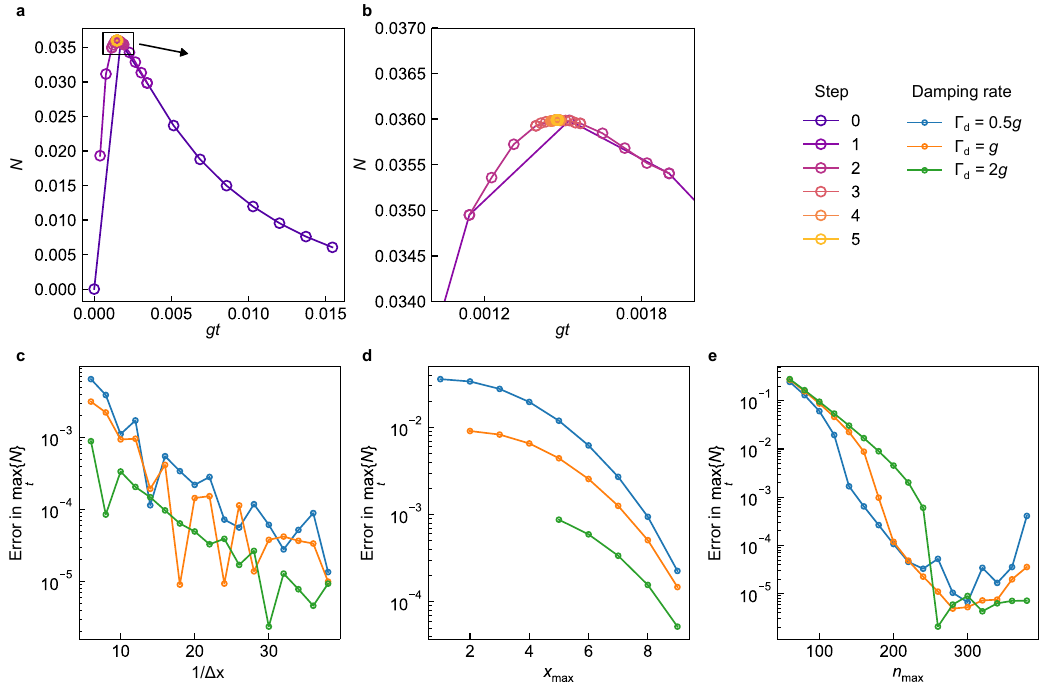}
  \caption{Procedure for computing the maximum negativity and convergence analysis.
    \textbf{a}, Illustration of consecutive refinement used to find the point of maximum negativity for a damping of $\Gamma_{\mathrm{s}}=0.5g$ and $s=6$. 
    \textbf{b}, Close-up of the point of maximum negativity in \textbf{a}.
    \textbf{c},\textbf{d},\textbf{e},
    Convergence of the error in the maximum negativity with respect to the grid resolution, $1/\Delta x$, the grid extent, $x_{\mathrm{max}}$, and the Fock space size, $n_{\mathrm{max}}$.
    For any given parameter, the two others are held fixed at their maximum value, namely $\Delta x=1/40$, $x_\text{max}=10$, and $n_{\text{max}}=400$.
    The error in ${\mathrm{max}}_t\{N\}$ for a given set of parameters is computed as the absolute difference between the computed quantities, ${\mathrm{max}}_t\{N\}$, for the given parameters and the maximally refined parameters.
    The error is shown for three different damping rates, $\Gamma_{\mathrm{s}}$, one of which corresponds to that used in \textbf{a} and \textbf{b}.
  }
    \label{fig:convergence}
\end{figure}

The Wigner functions displayed in Figs.~2a, 3a, and 4a of the main text are computed this way. To extract the maximum negativity as a function of time, the negativity is evaluated at a sequence of points which is refined around the maximum in a series of steps until the change in $N$ between neighboring points is below a tolerance of $10^{-5}$. This procedure is illustrated in Fig.~\ref{fig:convergence}a and b.
Convergence below $10^{-3}$ is independently verified for each of the three quantities $\Delta x$, $x_\text{max}$, and $n_{\text{max}}$, while keeping the other two fixed. This is shown in Fig.~\ref{fig:convergence}c, d, and e, respectively. 

\nocite{apsrev42Control} \putbib[references.bib,mainNotes.bib] 

\clearpage 
\end{bibunit}
\clearpage

\end{document}